\documentclass[12pt]{iopart}

\usepackage{iopams}  
\usepackage[numbers,sort&compress]{natbib}

\usepackage{graphicx}
\begin{document}

\title[]{Conformal Gradient Index Phononic Crystal Lenses: Theory and Application on Non-planar Structures}

\author{Hrishikesh Danawe}
\address{Department of Mechanical Engineering, University of Michigan, Ann Arbor, MI USA 48109}

\author{Serife Tol}
\address{Department of Mechanical Engineering, University of Michigan, Ann Arbor, MI USA 48109}
\ead{stol@umich.edu}
\vspace{10pt}

\begin{abstract}
The gradient index phononic crystal (GRIN-PC) lens concept has been proven very effective for focusing elastic waves at a desired location. Although well-studied for planar structures, GRIN-PC lenses for elastic wave focusing in curved structures are scarce and lack the theoretical framework for studying the wave focusing mechanism. In this work, we develop conformal GRIN-PC theory to analyze wave focusing in non-planar geometries and present a design framework for conformal GRIN-PC lenses to be implemented over curved structures. The proposed conformal GRIN-PC theory studies the wave propagation in a curved GRIN-PC lens using ray trajectories that meet at the focal spot of the lens. We apply the conformal GRIN-PC theory to accurately predict the focal region of the GRIN-PC lens implemented over a steel pipe and validate the results with numerical simulations. Further, the design framework is utilized to design a 3D-printed conical GRIN-PC lens. The elastic wave focusing in the conical lens is demonstrated using numerical simulations and is further validated with experiments.
\end{abstract}

%
\vspace{2pc}
\section{Introduction}

The gradient index (GRIN) lens concept is well studied in optics literature \cite{Moore80,Nishi86,Ohmi88,Koike94,GOMEZVARELA20121706,C005071G}, as it enables the creation of flat optical lenses overcoming the limitations of conventional spherical lenses in focusing light waves. The GRIN medium is composed of layered material of gradually varying refractive indices so that the light rays bend from the region of low refractive index towards the region of high refractive index. In the GRIN lens, the refractive indices of different layers are tailored to obtain a refractive index profile such that it results in the focusing of an incident beam of light at a desired location. According to GRIN optics theory \cite{GRINOptics}, a hyperbolic secant (HS) profile results in aberration-free focusing of meridional rays (i.e., rays propagating in planes that include the optical axis) for which the governing equations can be solved analytically to predict the focal spot. Thus, the HS profile is exceptionally used for designing GRIN lenses. 

With the emergence of phononic crystals (PCs), the concept of the GRIN lens was extended to acoustic or elastic waves using a layered structure called gradient index phononic crystal (GRIN-PC) lens. Phononic crystals are artificially engineered structures with spatially periodic structural features called scatterers that enable exceptional wave control due to Bragg scattering. By tailoring the geometric or material properties of the scatterer, the wave properties (such as wave speed) of the PCs can be altered to achieve unprecedented wave phenomena. In GRIN-PC lenses, the properties of scatterers in different layers are engineered such that the effective refractive index across the layers follows the HS profile transverse to the wave propagation direction, thus focusing an incident elastic or acoustic wave at the focal spot of the lens. The first GRIN-PC lens was designed using the GRIN optics theory to focus bulk acoustic waves in a planar 2D PC made of epoxy medium embedded with cylindrical rods as scatterers \cite{Lin2009}. The gradient HS profile of the effective refractive index was achieved by changing the diameter or material of the cylindrical rods in different layers or rows of the GRIN-PC lens. The follow-up studies on GRIN-PC lenses for elastic waves mainly considered focusing of symmetric (S0) and antisymmetric (A0) Lamb waves in plates at different length scales \cite{Wu2011,Zhao2012,CHIOU20141984,Jin2015,TolAPL,TOL2019AddManuf}. However, in most cases, discrepancies were observed between the numerical and theoretical focal distances of GRIN-PC lenses. The reason is that phononic crystals' anisotropic nature is not captured in the optical GRIN theory, which assumes PC as an isotropic medium of effective refractive index. To accurately predict the focal region of planar GRIN-PC lenses, Zhao et al. \cite{Zhao_2014} proposed an analytical ray tracing method utilizing the equal frequency contours (EFCs) of PC to locally determine the wave vector and group velocity in every row of the GRIN-PC lens. So, even if the GRIN-PC lenses are designed based on effective refractive indices to fit the HS profile, the wave-focusing mechanism can only be understood with EFCs. Recently, a ray theory was also proposed for wave propagation in more general, spatially graded, planar elastic metamaterials assuming local periodicity varying slowly in space compared to the unit cell length scale \cite{Dorn2022}. It also utilizes the locally computed wave vectors and group velocity vectors to trace the ray emanating from a point source. 

The GRIN-PC lenses are found very effective for the localization of wave energy benefiting many applications such as energy harvesting \cite{TolAPL,TOL2019AddManuf}. Although well studied for planar structures, their application on curved structures was not yet explored before we demonstrated the first conformal GRIN-PC lens for focusing Lamb waves in pipe-like structures \cite{Danawe2020APL,Danawe2020SPIE}. The cylindrical GRIN-PC lens was made of steel stubs attached to the outer surface of the steel pipe, and the effective HS refractive index profile was achieved by tailoring the stub heights around the circumference of the pipe. The lens was found very effective for multi-mode broadband wave focusing of ultrasonic guided waves in pipes \cite{DanaweIDETC}. However, similar to planar GRIN-PC lenses, we found discrepancies in focal distances determined from optical GRIN theory and numerical simulations because of the anisotropy of PCs. Thus, there is a need for the development of conformal GRIN-PC lens theory to understand the focusing mechanism in non-planar structures.

 In this work, we propose a conformal GRIN-PC theory for tracing ray trajectories inside a curved GRIN-PC lens. We adopt the analytical approach of calculating the beam path inside a planar GRIN-PC lens presented by Zhao et al. \cite{Zhao_2014} and apply it to a more general non-planar geometry via coordinate transformation. The theory is applied to accurately determine the focal region of the cylindrical GRIN-PC lens previously implemented over steel pipe \cite{Danawe2020APL,Danawe2020SPIE}. Using the proposed theory, we further design a 3D-printed conical GRIN-PC lens and demonstrate multi-mode elastic wave focusing of guided Lamb waves in conical structures commonly found in civil, mechanical, and aerospace industries. The 3D-printed GRIN-PC lens is numerically and experimentally tested for elastic wave focusing. The presented design framework is crucial to extend the concept of GRIN-PC lenses for elastic wave focusing beyond planar structures that can benefit many applications including nondestructive testing, sensing, energy harvesting, etc.
 
\section{Conformal GRIN-PC Theory}

A GRIN lens with hyperbolic secant distribution of refractive index results in focusing of normally incident beam at a distance of $\pi/2\alpha$ according to the optical GRIN theory, which is conventionally adopted for acoustic/elastic waves. The optical (a.k.a. conventional) GRIN theory assumes perfectly circular EFCs which means the wavevector magnitude and group speed are the same in all directions. This results in focusing the normally incident beam on the lens at a single location. However, in general, the EFCs of phononic crystals are not perfectly circular. Thus, tracing the ray trajectory in the GRIN-PC lens requires computing the EFCs in each row to account for the anisotropy. Zhao et al. \cite{Zhao_2014} presented a framework for tracing a ray path in a planar GRIN-PC lens by locally computing the wave vectors and group velocity vectors. The ray path across the neighboring unit cell layers was determined using the combination of Snell's law and Poynting vector (i.e., group velocity vector). We utilize a similar approach for tracing the ray trajectories in a non-planar GRIN-PC lens with a coordinate transformation from cartesian to cylindrical coordinates. The wave vectors and group velocity vectors are determined from the EFCs of curved unit cells. To demonstrate the computation of ray trajectories in non-planar structures, we take the example of a cylindrical GRIN-PC lens implemented over steel pipe \cite{Danawe2020APL}, as shown in Fig. \ref{fig:NP_Ray}(a). In the cylindrical coordinate system, the wave vector has two non-zero components: $k_x$ along the pipe axis and $k_{\phi}$ along the pipe circumference. The wavevector components are different in different rows of the GRIN-PC lens due to the gradient distribution of stub heights. Thus, the wavevector magnitude is a function of angular distance $\phi$ from the centerline and the wave propagation direction $\theta$ measured with respect to $x$-direction, as shown in the EFC plots in Fig. \ref{fig:NP_Ray}(b). The wavevector components are thus given as:\begin{figure}[t]
\centering
\includegraphics[width=0.9\linewidth]{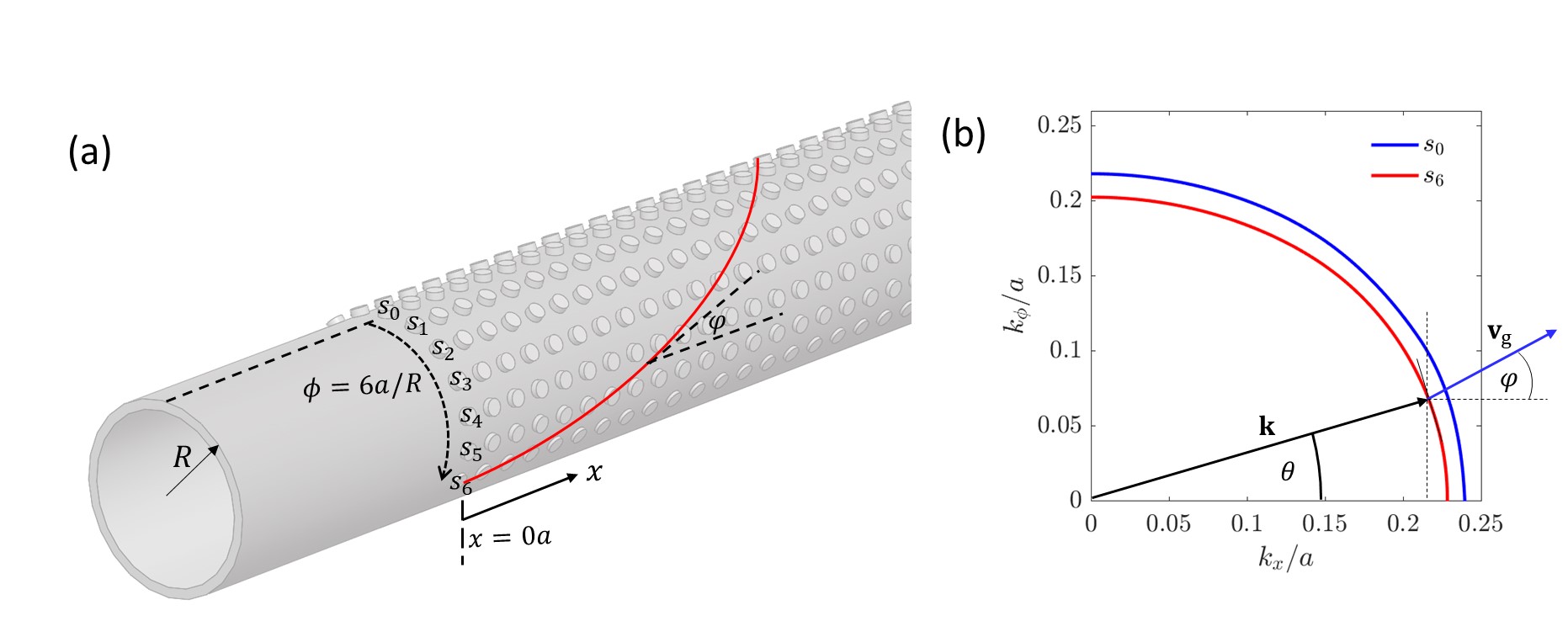}
\caption{(a) The ray trajectory in cylindrical GRIN-PC lens.  (b) EFCs for two different rows of GRIN-PC lens. $\textbf k$ is the wavevector at row $s_0$ whose x-component is equal to the wavevector at row $s_6$. $\textbf v_g$ is the group velocity vector.  }
\label{fig:NP_Ray}
\end{figure} 
\begin{equation}
\label{wavevectorcomp}
    k_x=k(\phi,\theta) \cos(\theta),  k_{\phi}=k(\phi,\theta) \sin(\theta)
\end{equation}
Now, because of the anisotropy the group velocity vector defined as $\textbf v_g=\nabla_k\omega(\textbf k)$ is at an angle $\varphi$ to the wave vector such that the slope of ray trajectory in curved GRIN-PC lens is given by:
\begin{equation}
\label{eq:trajectoryslope}
    \tan(\varphi)=-\frac{\partial k_x}{\partial\theta} \left(\frac{\partial k_{\phi}}{\partial\theta}\right)^{-1}
\end{equation}
If we consider a beam normally incident on the GRIN-PC lens, Snell's law states that the $x$-component of wavevector $k_x$ is conserved across the interface of two consecutive rows of the GRIN-PC lens. Thus, the wavevector tilts gradually from a horizontal position ($\theta=0$) at the beginning of the lens to attain maximum angle with respect to the $x$-axis at the centerline. The ray tracing starts at the beginning of each row ($x=0a,\phi=na/R$), where $n\in [1,6]$ is the row number, and  $k(\phi=na/R,\theta=0)=k_x$ is the initial wavevector which is conserved due to the Snell's law. Now, to predict the ray path, we move closer to the centerline in incremental steps of angular distance $\textrm d\phi$ to search for the wave vector at $\phi=na/R-m\textrm d\phi$, where $m$ is the step number, such that $k(\phi=na/R-m\textrm d\phi,\theta)=k_x$, i.e., we find the unknown angle $\theta$ with the help of EFCs. Next, we determine the slope of the ray trajectory, $\tan(\varphi)$ using equation \ref{eq:trajectoryslope}. The axial location $x$ for the ray at angular position $\phi=na/R-md\phi$ is then determined using the following iterative relation:

\begin{equation}
\label{xtrajectory}
   x(\phi=na/R-m\textrm d\phi)=x(\phi=na/R-(m-1)\textrm d\phi)+\frac{R\textrm d\phi}{\tan(\varphi)}
\end{equation}
Note that even if the GRIN-PC lens is divided into discrete rows with gradually varying stub heights, the incremental step $\textrm d\phi$ is chosen much smaller than the angular stub spacing to obtain converging results. Since the EFCs are only calculated for the unit cells in discrete rows, the ray trajectory calculation at locations in between two consecutive unit cell rows is done using interpolated EFCs by assuming a continuous variation of stub height.

\section{Ray Trajectories in the Curved GRIN-PC Lens}

\begin{figure}[ht]
\centering
\includegraphics[width=0.9\linewidth]{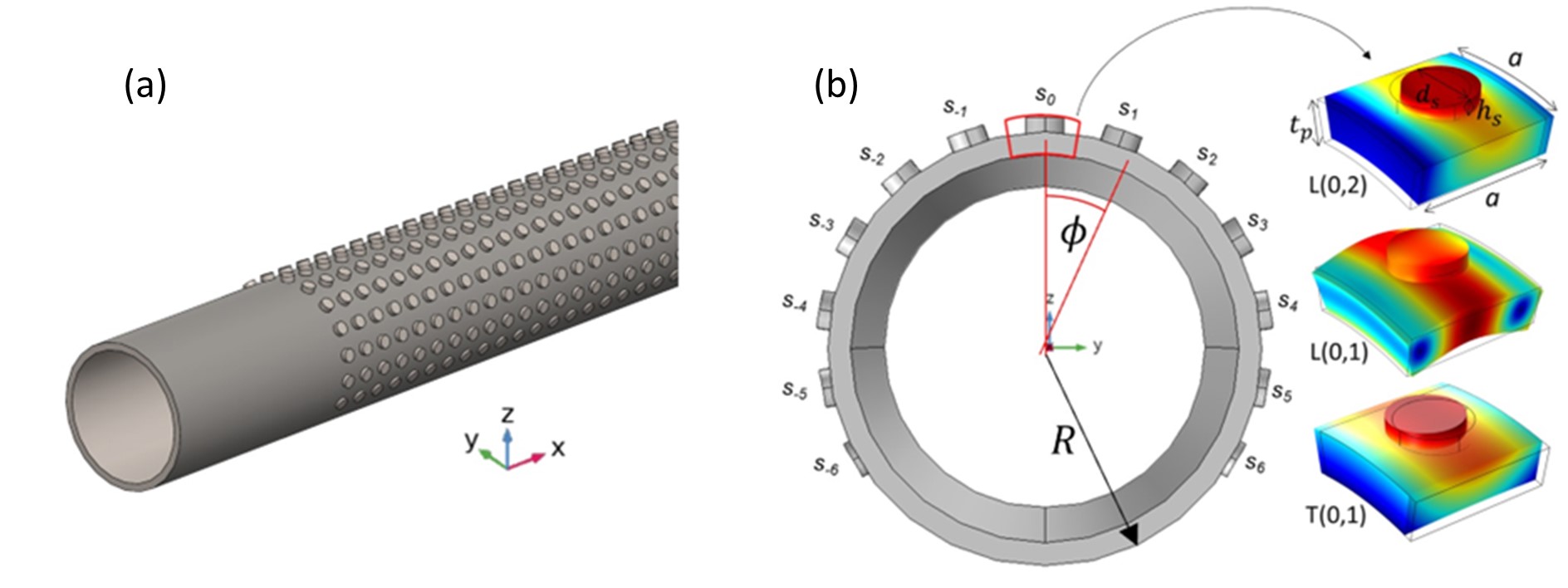}
\caption{(a) Pipe integrated with a curved GRIN-PC lens. (b) A GRIN-PC lens design for steel pipe consisting of externally attached steel stubs of varying heights. Mode shapes of the unit cell for the three fundamental pipe modes are shown in the inset.}
\label{fig:Design_pipe}
\end{figure}

A curved GRIN-PC lens integrated with steel pipe is depicted in Fig. \ref{fig:Design_pipe}(a) \cite{Danawe2020SPIE}. It consists of steel pipe with outer radius $R=$57.15 mm with externally attached steel stubs of constant diameter $d_s=$10 mm and varying height across the circumference, as shown in Fig. \ref{fig:Design_pipe}(b). The stubs are uniformly spaced in the axial and circumferential direction such that the inter-stub distance equals the unit cell length. The unit cell mode shapes are depicted in the inset for three fundamental pipe modes, L(0,2), L(0,1), and T(0,1). The unit cell length is $a=20$ mm and the pipe wall thickness is $t_p=6$ mm. The stub heights are tailored to obtain a hyperbolic secant (HS) profile of the refractive index.  The stub height is maximum at the centerline, $S_0$, and decreases symmetrically on either side of the centerline  up to the lens edges, $S_{\pm6}$. The height profile is obtained as (4.5000, 4.4646, 4.3188, 4.0668, 3.6588, 2.9682, 1.9158) mm at locations $S_0$ to $S_{\pm6}$, respectively. In order to trace ray trajectories, we define a sector angle $\phi$ measured from the centerline such that  $\phi R/a$ is 1 at location $S_1$, 2 at location $S_2$, and so on, in the cylindrical coordinate system. 
\begin{figure}[ht]
\centering
\includegraphics[width=0.95\linewidth]{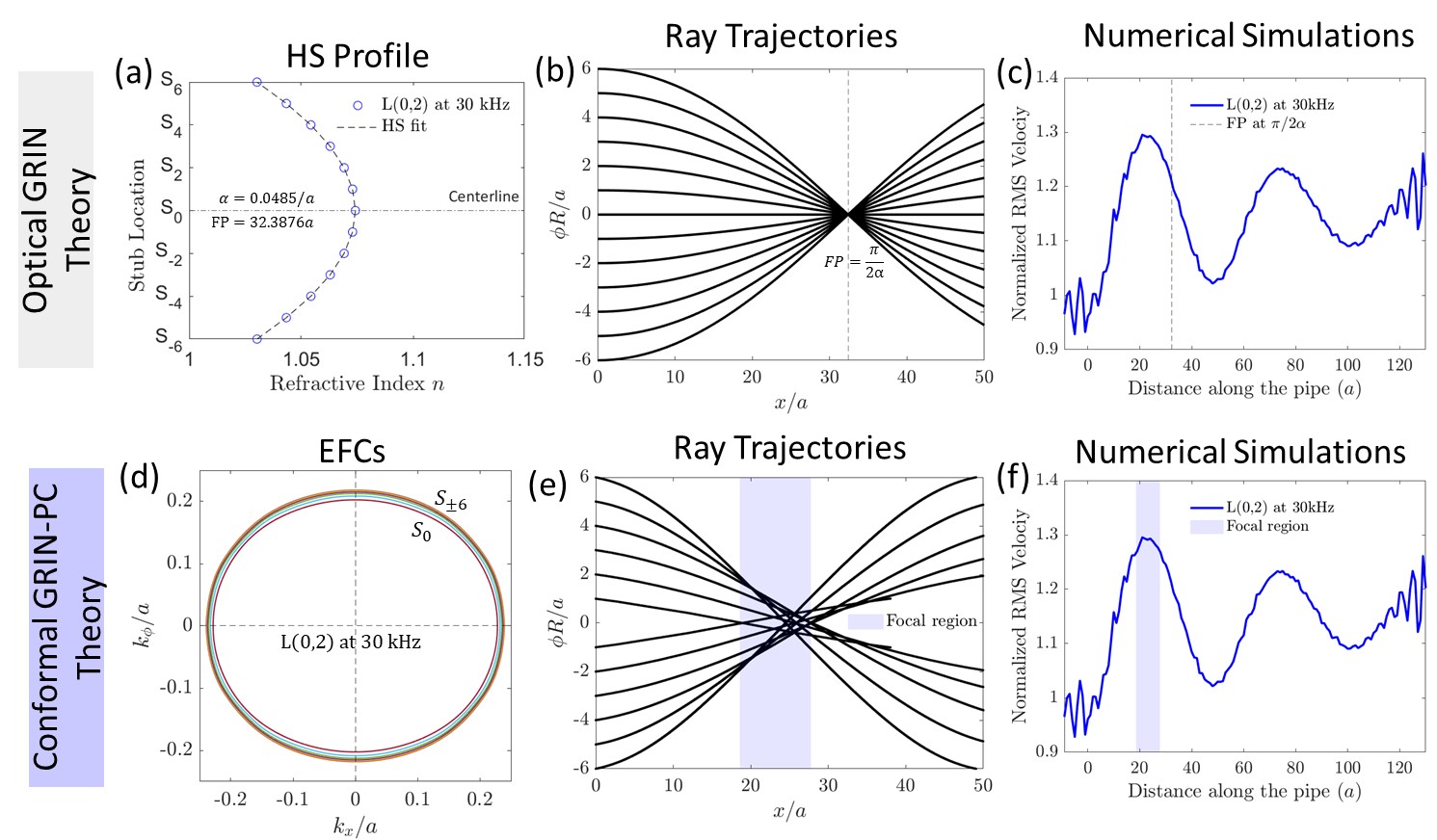}
\caption{Comparison of ray trajectories for L(0,2) mode at 30 kHz derived from conventional and non-planar GRIN theories. (a) Hyperbolic secant (HS) profile of refractive index. (b) Ray trajectories using optical GRIN theory. (c) Focal point of $\pi/2\alpha$ predicted using optical GRIN theory compared with numerical simulations. (d) Equal frequency contours for different unit cell rows of the GRIN-PC lens. (e) Ray trajectories using conformal GRIN-PC theory. The shaded region is the predicted focal region based on the intersection of the ray trajectories at the centerline. (f) The focal region predicted using conformal GRIN-PC theory compared with numerical simulations.}
\label{fig:NP_RT}
\end{figure}

An HS profile is well studied in GRIN optics literature for aberration-free focusing as parallel rays meet at a single point after being gradually refracted through a GRIN medium. The hyperbolic secant profile for pipe is defined as  $n=n_{0}\textrm{sech}(\alpha \phi R)$, where $n_0$ is the refractive index at the lens centerline and $\alpha$ is the gradient coefficient. The refractive index distribution of the GRIN-PC lens is obtained from the dispersion variation of the unit cell for different stub heights. The refractive index distribution for L(0,2) mode at 30 kHz is fitted with HS profile as shown in Fig. \ref{fig:NP_RT}(a), for which the ray trajectories meet at a distance of $\pi/2\alpha$ according to the conventional GRIN theory, as shown in Fig. \ref{fig:NP_RT}(b). The rays bend from the region of low refractive index at the edges towards the region of high refractive index at the centerline. We have previously studied the focusing effect of curved GRIN-PC lenses via time-domain numerical simulation in COMSOLMultiphysics \cite{DanaweIDETC}. To determine the focal region, we obtained RMS velocity plots along the lens centerline normalized with RMS velocity in baseline (i.e., pipe without GRIN-PC lens) as depicted in Fig.\ref{fig:NP_RT}(c) for the L(0,2) mode at 30 kHz. The normalized velocity amplitude is close to 1 before the lens start and increases along the pipe length to attain maximum value at the first focal point. The velocity amplitude decreases past the first focal point and peaks again at the second focal point because of refocusing. The focal region is identified with the maximum velocity amplitude and compared with the focal point obtained using optical GRIN theory. The predicted first focal point using optical GRIN theory $FP=32.3876a$ lies beyond the highest intensity point in numerical simulations. As previously explained in the introduction, the discrepancy is because the phononic crystals are not generally isotropic. The anisotropy of PCs is captured with equal frequency contours (EFCs) obtained from unit cell simulations. The EFCs of L(0,2) mode at 30 kHz are shown in Fig. \ref{fig:NP_RT}(d). The equal frequency contours are not perfect circles, meaning that the wave vectors and wave speeds are different along different directions. Hence, a single value of the refractive index in the HS profile, obtained by averaging it in different directions, could not predict the focal point accurately using the conventional GRIN theory. We implement a conformal GRIN-PC theory for accurately predicting the focal region of GRIN-PC lenses by utilizing EFCs. The conformal GRIN-PC theory utilizes directional phase speeds and group velocities to predict the path a ray would take in a GRIN-PC lens. The directional phase and group velocities are obtained from EFCs at every single location on the ray trajectory. The ray trajectories calculated using conformal GRIN-PC theory for L(0,2) mode at 30 kHz are depicted in Fig. \ref{fig:NP_RT}(e). The ray trajectories do not meet at a single location as previously predicted by the conventional GRIN theory. The focal region of the GRIN-PC lens is determined from the intersection of ray trajectories at the lens centerline ($\phi=0$). The predicted focal region from conformal GRIN-PC theory matches exactly with the highest intensity region at the first focal point in numerical simulations, as shown in Fig. \ref{fig:NP_RT}(f).

\section{GRIN-PC Lens Implementation in Conical Structures}
\begin{figure}[ht]
\centering
\includegraphics[width=1\linewidth]{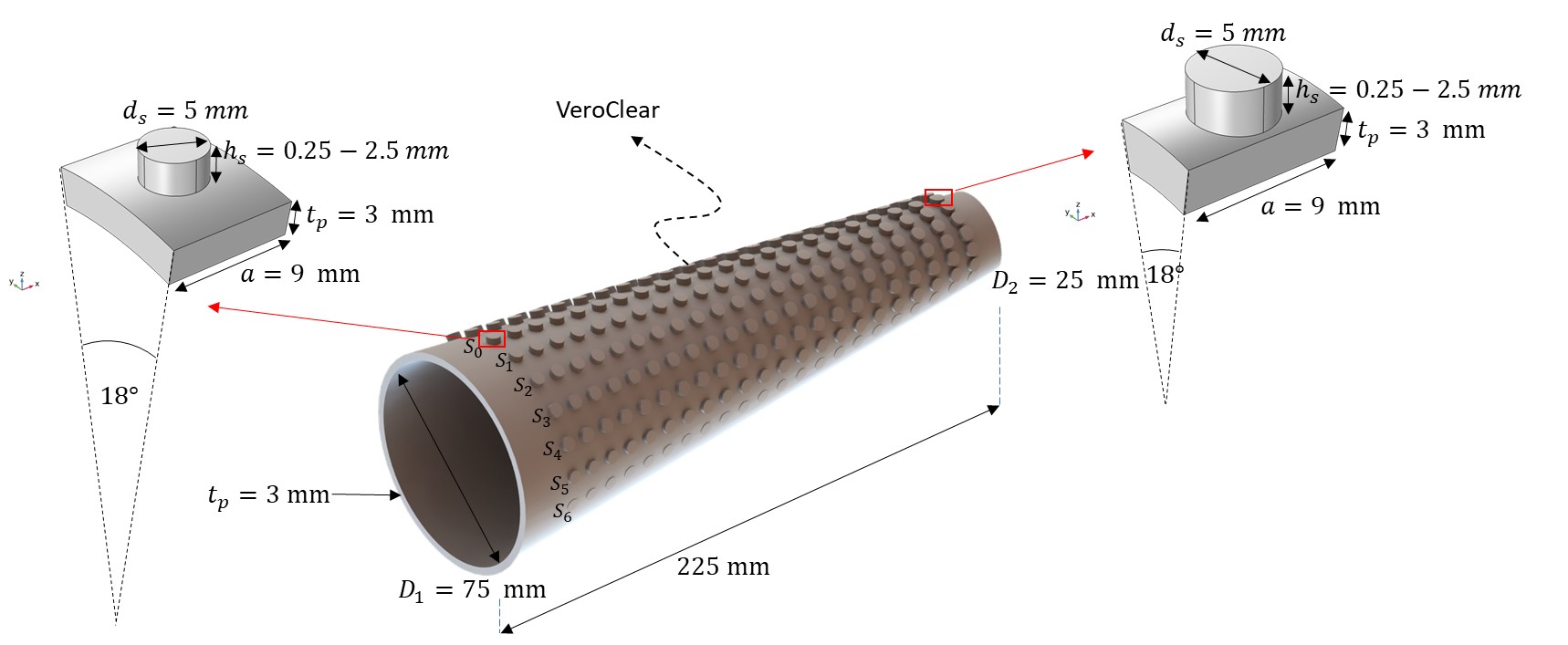}
\caption{Conical GRIN-PC lens design consisting of externally attached cylindrical stubs on the cone surface. The insets show the unit cells at the GRIN-PC lens's start and end.      }
\label{fig:Design}
\end{figure}

To demonstrate the applicability and effectiveness of the proposed theory for any curved structure, we consider a conical shell of uniform wall thickness. Conical shells are commonly found in civil, mechanical, and aerospace industries, which require structural health monitoring and can also serve as a platform for enhanced energy harvesting of ambient structural vibrations via guided wave focusing. We chose to design a conformal GRIN-PC lens for the conical structure similar to the steel pipe with externally attached cylindrical stubs of varying heights on the outer surface. The GRIN-PC lens integrated with a conical structure is depicted in Fig. \ref{fig:Design}. The conical GRIN-PC lens comprises stubbed unit cells, representing a phononic crystal pipe of an infinite extent. The wave propagation characteristics of the phononic crystal pipe are obtained by applying periodic Floquet boundary conditions at the unit cell sides and solving for the eigenfrequency solutions by sweeping wave vectors in the first Brillouin zone. We compute the dispersion curves of the unit cell in COMSOL Multiphysics using solid mechanics physics and eigenfrequency study. The unit cell is made of VeroClear with material properties: $\rho=1170$ kg/m$^{3}$, $E=2.55$ GPa, $\nu=0.3$. The Floquet periodicity boundary conditions in COMSOL are as follows:
\begin{equation}
\label{PeriodicBC}
    \textbf u_{dst}=\textbf u_{src}\cdot e^{i \textbf k \cdot (\textbf r_{dst}-\textbf r_{src})}
\end{equation}
where,  $\textbf u_{src}$ and $\textbf u_{dst}$ are displacement vectors at the source and destination boundaries of the unit cell, respectively. Similarly, $\textbf r_{src}$ and $\textbf r_{dst}$ are position vectors at the source and destination boundaries of the unit cell, respectively, and $\textbf k$ is the wave vector. The cone, along with the GRIN-PC lens, is made of VeroClear, which is a 3D printable polymer available with PolyJet printers. The prototype cone is 225 mm long with wall thickness $t_p=3$ mm, and the internal diameter of the cone varies from $D_1=75$ mm at one end to $D_2=25$ mm at the other end. The axial and angular spacing between the stubs of the GRIN-PC lens is kept constant throughout the lens to respect the geometry of the cone for guided wave propagation. The axial and angular spacing between the neighboring stubs is $9$ mm and 18$^{\circ}$, respectively. The GRIN-PC lens is 22 unit cells long in the axial direction and has 13 unit cell rows along its circumference. The axial length of the unit cell equals $a=9$ mm, and the stub diameter equals $d_s=5$ mm. The circumferential length of the unit cell varies along the cone axis due to varying diameter and constant angular spacing of 18$^{\circ}$. The stub heights are tailored in the circumferential direction to realize hyperbolic secant (HS) refractive index distribution.  The stub height is maximum at the centerline unit cell row $S_0$ and minimum at the edge rows of the lens $S_{\pm6}$. The stub height profile is kept constant along the axis of the cone. We compute dispersion curves of unit cells for different stub heights as previously done for steel pipe. However, since the curvature of the cone varies along its axis, the angular length of the unit cell is different at every location along the axis. Hence, the dispersion curves are computed not only for different stub heights but also for different axial locations. 

\begin{figure}[ht]
\centering
\includegraphics[width=0.9\linewidth]{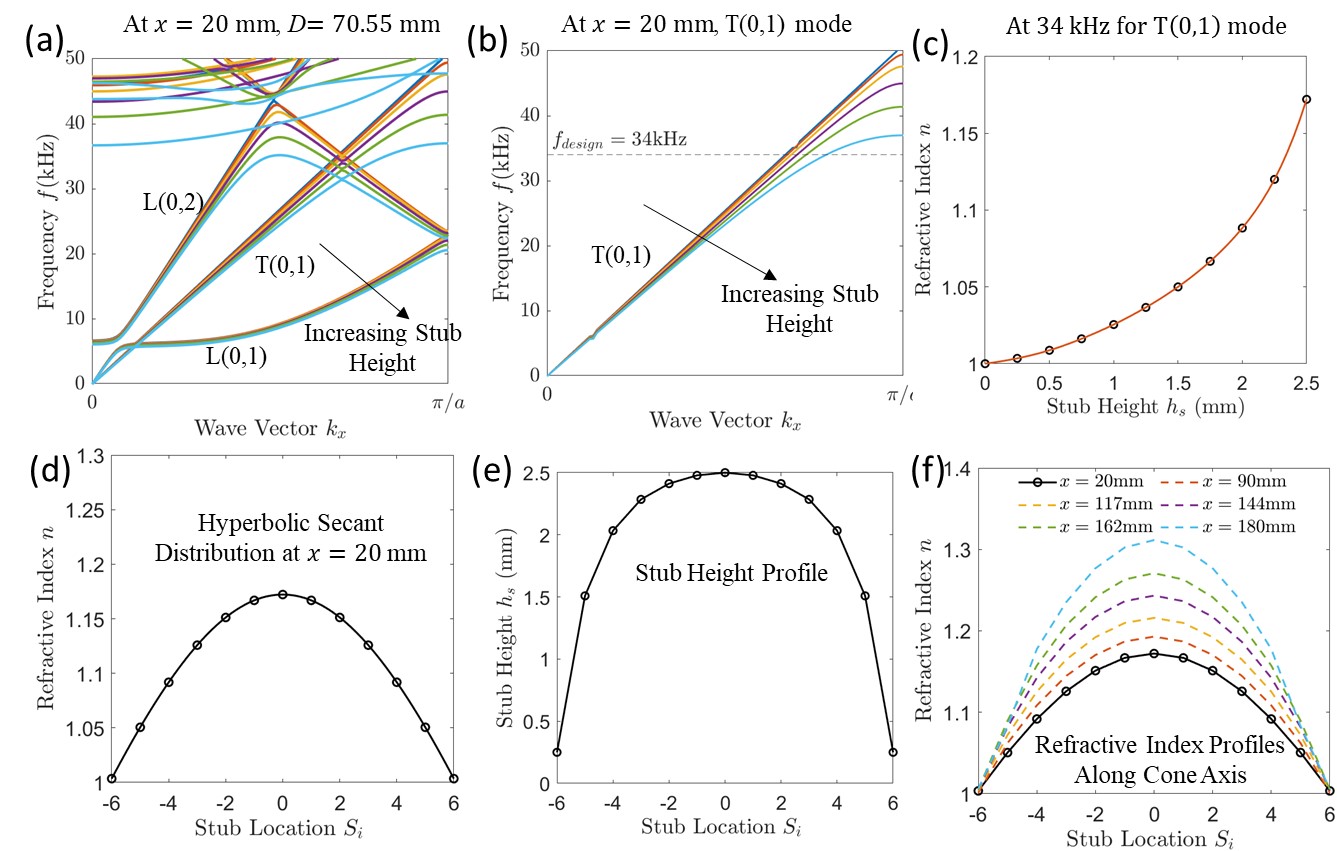}
\caption{(a) Dispersion curves of a phononic crystal pipe with internal diameter $D=70.55$ mm locally calculated at axial location $x=20$ mm along the cone axis. (b) Dispersion variation of T(0,1) mode at $x=20$ mm. (c) Refractive index of T(0,1) mode as a function of stub height at $x=20$ mm and $f_{design}=34$ kHz. (d) Hyperbolic secant (HS) refractive index distribution for T(0,1) mode at $x=20$ mm and $f_{design}=34$ kHz based on maximum and minimum stub heights. (e) Stub height profile for conical GRIN-PC lens. (f) Variation in refractive index distribution of T(0,1) mode at $f_{design}=34$ kHz along the axis of the cone.}
\label{fig:GRIN_Design}
\end{figure}

The dispersion curves of the unit cell at an axial distance of $x=20$ mm ($x=0$ mm is the left end of the cone with internal diameter $D_1=75$ mm) are shown in Fig. \ref{fig:GRIN_Design}(a) for stub heights ranging from 0.25 mm to 2.5 mm. The dispersion curves represent three fundamental pipe modes (L(0,2), L(0,1), and T(01)) of a pipe with internal diameter $D=70.77$ mm and wall thickness $t_p=3$ mm. Note that the dispersion curves are locally calculated considering that the unit cell represents a phononic crystal pipe of diameter equaling the cone diameter at that location. As expected, the dispersion curves shift to a lower frequency with increasing stub height. We found that the torsional T(0,1) mode dispersion curve does not change with the pipe diameter for plain pipe, and thus it propagates with the same wave speed throughout the cone for a given excitation frequency. However, the wavelengths of longitudinal L(0,1) and L(0,2) modes are affected by the diameter, and thus their propagation speed changes along the cone axis. Thus, for simplicity, we chose T(0,1) mode for designing the GRIN-PC lens whose dispersion variation is shown in Fig. \ref{fig:GRIN_Design}(b). The design frequency is chosen just below the Bragg bandgap for T(0,1) mode at 34 kHz corresponding to the maximum stub height of 2.5 mm. From dispersion variation, we obtained the refractive index $n=v/v_{\Gamma X}$ as a function of stub height where $v$ is the phase velocity of T(0,1) mode in the plain pipe and $v_{\Gamma X}$ is the phase velocity of T(0,1) mode in phononic crystal pipe. The refractive index as a function of stub height is plotted in Fig. \ref{fig:GRIN_Design}(c). The refractive index increases with increasing stub height, indicating that the wave speed is slower for higher stub heights. Thus, the wave travels faster at the edges of the lens where the stub height is minimum and it travels slower at the centerline where the stub height is maximum. The stub height profile around the circumference of the cone is depicted in Fig. \ref{fig:GRIN_Design}(e), which is obtained to follow the HS profile of refractive index, as shown in Fig. \ref{fig:GRIN_Design}(d). The gradient coefficient equals $\alpha=0.0953/a$ for the HS distribution of T(0,1) mode at $x=20$ mm and $f_{design}=34$ kHz for which the first focal point is predicted at $16.5a$ according to optical GRIN theory. Now, even if the dispersion curves of plain pipe remain the same for different diameters, the unit cell with stubs shows variation in dispersion curves of T(0,1) mode as the diameter varies. Thus, for the same stub height profile, the refractive index distribution changes along the cone axis, as depicted in Fig. \ref{fig:GRIN_Design}(e). Therefore, conventional GRIN theory is insufficient to predict the focal point of the GRIN-PC lens for a conical structure. Hence, we numerically investigate the focusing of the three pipe modes using the designed GRIN-PC lens for the cone in the next section.  

\subsection{Numerical Results}

\begin{figure}[t]
\centering
\includegraphics[width=0.9\linewidth]{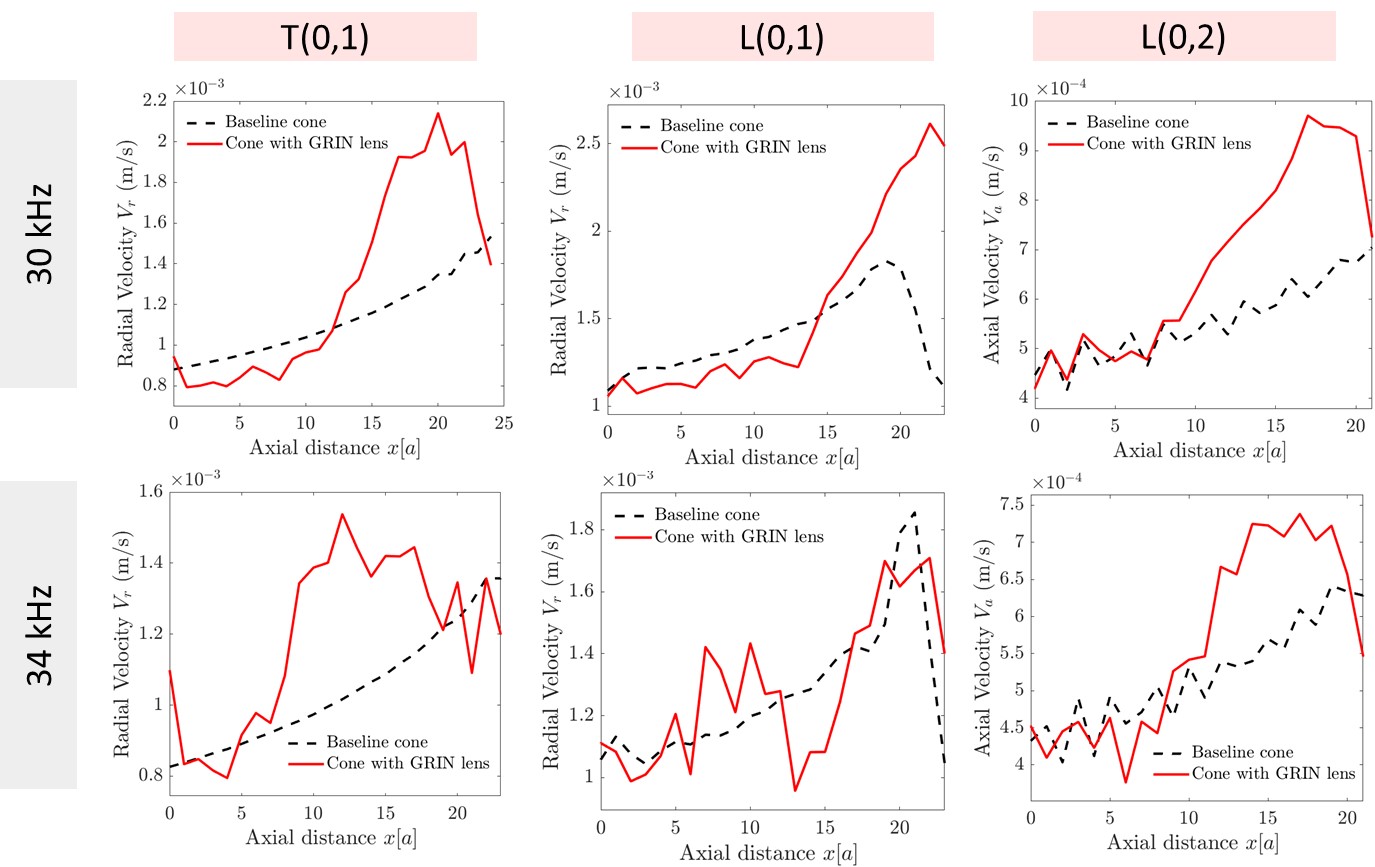}
\caption{The RMS velocity plots along the lens centerline show amplification of velocity amplitude with the GRIN lens compared to the baseline cone. The focusing results are obtained for three pipe modes T(0,1), L(0,1), and L(0,2) at two different frequencies of 30 kHz and 34 kHz.}
\label{fig:Sim_Results}
\end{figure}

The conical GRIN-PC lens design was numerically tested for multimode wave focusing through time-domain numerical simulations. The simulation model consists of a cone integrated with a GRIN-PC lens made of VeroClear, as shown in Fig. \ref{fig:Design}. Using solid mechanics physics, the time domain numerical simulations on conical GRIN-PC lens were run in COMSOL Multiphysics. The CAD model of the cone integrated with the GRIN-PC lens (see Fig. \ref{fig:Design}) was built in Solidworks and imported into COMSOL for finite element simulations. The material of the cone is VeroClear which was modeled using linear elastic solid. Low-reflecting boundary conditions were applied at the two ends of the cone to avoid reflected waves interfering in the lens region. A 7-cycle sine burst excitation was applied at the left edge of the cone where the inner diameter is $D_1=75$ mm. The edge load was applied in tangential, radial, and axial directions for exciting T(0,1), L(0,1), and L(0,2) modes, respectively. The finite element model has tetrahedral mesh elements with a maximum element size of $\lambda/20$, where $\lambda$ is the wavelength of the excited mode. The time-dependent study was run with sufficiently smaller time steps to obtain a converging solution. The RMS velocity was extracted at the centerline unit cell row $S_0$ and it was compared with the cone without the GRIN-PC lens, as shown in Fig. \ref{fig:Sim_Results}. 

The plane wave excited at the left end of the cone starts to bend towards the lens centerline as it propagates through the lens. This is because the refractive index is highest at the lens centerline and gradually decreases towards the lens edges. Thus, the plane wave travels faster at the edges and slower at the centerline resulting in the bending of the wavefront from the region of low refractive index to the region of high refractive index. Figure \ref{fig:Sim_Results} shows RMS velocity at the lens centerline compared to the baseline cone for three different pipe modes at the design frequency of 34 kHz. Note that the lens starts at $x=0a$ and ends at $x=22a$ along the cone axis. The focusing results are also obtained at a frequency of 30 kHz away from the design frequency to demonstrate the broadband operation of the lens. Note that the velocity amplitude increases gradually in the baseline cone along its axis because of decreasing circumference. With the GRIN-PC lens, the velocity curve peaks above the baseline velocity curve at the focal point of the lens because of the focusing effect. The focal point location is different for different modes at different frequencies. At 30 kHz, all the modes focus toward the end of the lens. The focal points at the design frequency of 34 kHz are closer than at 30 kHz. The maximum amplification of velocity amplitude for T(0,1) mode at the design frequency of 34 kHz is obtained at a distance equal to $12a$, which is shorter than the predicted focal length of $16.5a$. As stated earlier, this is partly attributed to the changing refractive index distribution along the cone axis, as shown in Fig. \ref{fig:GRIN_Design}(f), and partly because of the anisotropy of the phononic crystal, which is not accounted for in the conventional GRIN theory. Nonetheless, broadband multimode focusing is numerically demonstrated with maximum amplification factors of 1.59, 2.19, and 1.56 at 30 kHz and 1.51, 1.20, 1.33 at 34 kHz for T(0,1), L(0,1), and L(0,2) modes, respectively. 

\subsection{Experimental Validation}
\begin{figure}[ht]
\centering
\includegraphics[width=\linewidth]{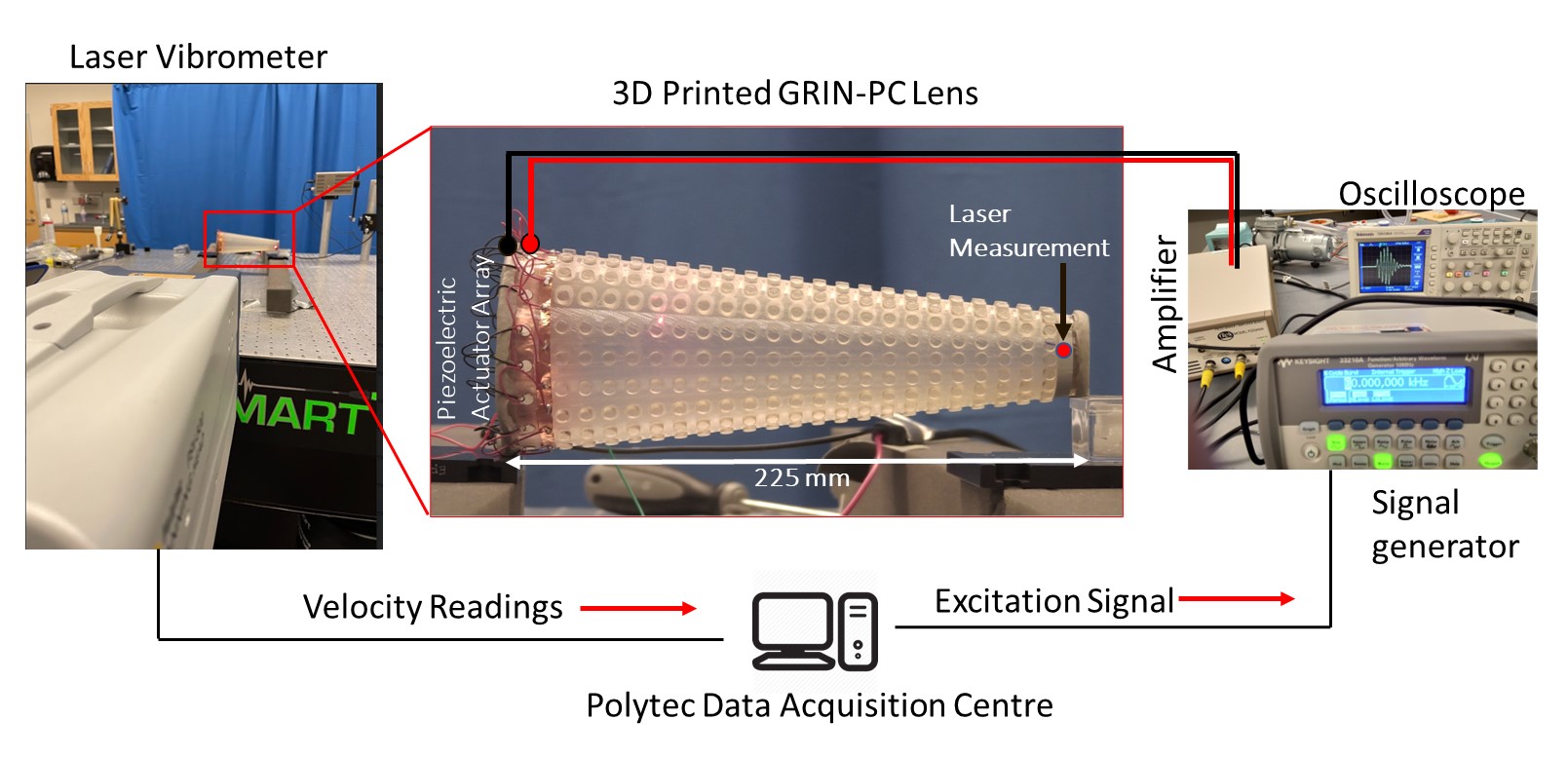}
\caption{(e)xperimental setup for vibration measurement of GRIN-PC lens integrated with 3D-printed cone.}
\label{fig:Exp_Setup}
\end{figure}

We further validate the wave focusing ability of conical GRIN-PC lens through laboratory experiments using a Polytec laser vibrometer and data acquisition system. The experimental setup is depicted in Fig. \ref{fig:Exp_Setup}. The cone integrated with GRIN-PC lens was 3D-printed using Stratasys J750 Polyjet 3D printer using VeroClear material, which is a rigid transparent polymer that simulates PMMA (polymethyl methacrylate). The cone is 225 mm long and has other dimensions similar to the numerical model, as depicted in Fig. \ref{fig:Design}. The cone was supported at both ends using soft supports placed on the vibration isolation table. An absorbing clay was applied at both ends of the cone to reduce wave reflections. An array of piezoelectric actuator disks of diameter equal to 5 mm and thickness of 0.4 mm were glued on the cone surface around its circumference with a layer of copper tape in between. The copper tape provides electrical contact to the bottom electrodes of the actuators, whereas the free top surface serves as the other electrode. The actuators vibrate radially to excite longitudinal plane waves in the cone right before the lens starts. The actuator array was excited using a signal generator connected to a power amplifier. The out-of-plane velocity signal was measured on the cone surface at the end of the GRIN-PC lens, where focusing of longitudinal modes is expected at 30 kHz from numerical simulations. The time domain velocity signal measured on the cone surface was stored in the Polytec data acquisition center. The experiments were conducted on a 3D-printed cone integrated with GRIN-PC lens using a setup consisting of a vibration isolation table, Polytec PSV 500 laser vibrometer, a Keysight 33210A function generator, TReK PZD350A amplifier, and data acquisition system, as shown in Fig. \ref{fig:Exp_Setup}. The function generator generates a 5-cycle sine burst signal with a peak-to-peak amplitude of 1V and signal time of 800$ \mu$sec. The time delay between two consecutive bursts was set to 50 msec. The burst signal was amplified by the TReK amplifier before supplying it to the piezoelectric actuator disks from Steminc Inc (PZT-4, radial mode vibration). The laser vibrometer was set to measure out-of-plane velocity on the cone surface with a sampling frequency of 0.625 MHz with 10-time averages. The vibrometer was in-sync with the signal generator, and the measured velocity data was stored in the data acquisition center.

\begin{figure}[ht]
\centering
\includegraphics[width=0.8\linewidth]{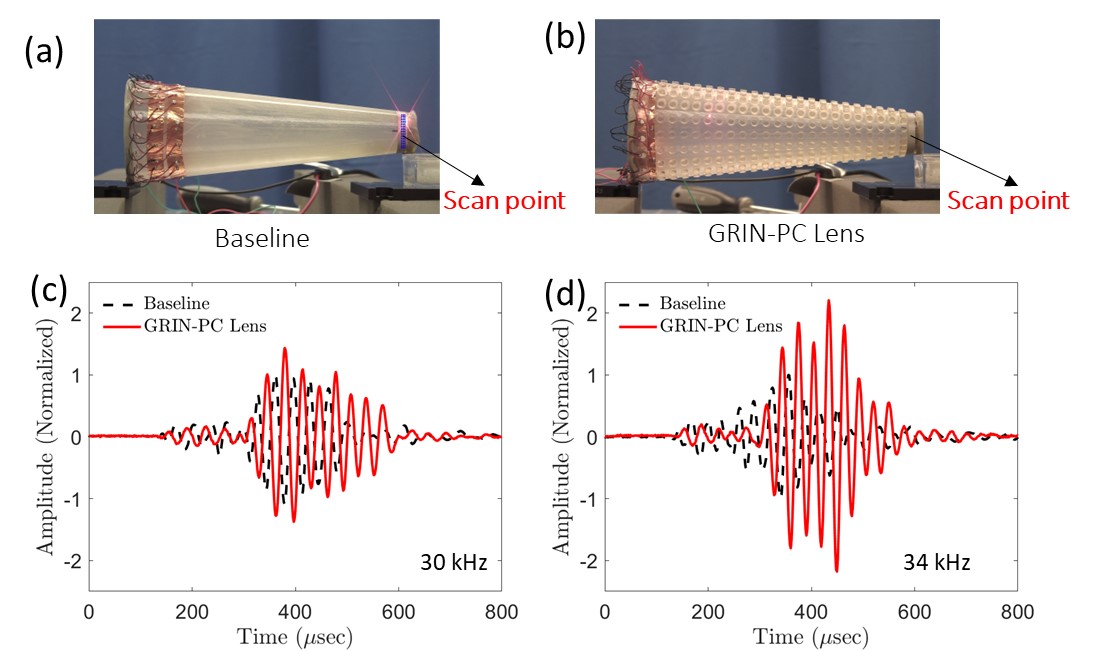}
\caption{(a) Baseline cone 3D-printed using VeroClear. (b) Cone integrated with GRIN-PC lens 3D-printed using VeroClear. (c) Velocity signal measured on the cone surface at 30 kHz. (d) Velocity signal measured on the cone surface at 34 kHz.  }
\label{fig:Exp_Results}
\end{figure}

The velocity signals captured using the laser vibrometer for the baseline cone depicted in Fig.\ref{fig:Exp_Results}(a) and the cone integrated with the GRIN-PC lens depicted in Fig.\ref{fig:Exp_Results}(b) are plotted in Fig. \ref{fig:Exp_Results}(c) and (d)  at frequencies 30 and 34 kHz, respectively. The piezoelectric actuator array excites only the longitudinal modes and the laser only measures out-of-plane velocity; thus, the waveform captured using the vibrometer corresponds to L(0,1) mode. The velocity signal is amplified at both the excitation frequencies due to the focusing effect of the GRIN lens. The maximum velocity amplitude with the GRIN-PC lens is about two times higher than that of the baseline cone at 34 kHz. The amplification at 30 kHz is not significant at the measured location. Note that the numerical results predict about two times amplification of velocity amplitude at 30 kHz, which is instead observed at 34 kHz in experiments. This shift in the frequency might be the because of the uncertainty in material properties of 3D-printed polymers as documented in the literature\cite{Barclift2012}. Several aspects, such as UV light exposure while printing affect the properties of 3D-printed materials. In fact, the material properties vary over a large range and are strongly affected by the printing process, as reported in previous studies. Stratasys has specified the range of Young's modulus for the VeroClear material, which is in between 2 GPa and 3GPA. However, the reported values in some studies for similar 3D-printed polymers go beyond this range\cite{Barclift2012}. In numerical simulations, Young's modulus of VeroClear is chosen as 2.55 GPa, which might differ from the actual material properties of 3D-printed samples in experiments. Also, the 3D-printed samples of the cone are printed layer by layer, which results in anisotropic behavior and cannot be accounted for in numerical simulations. Despite these uncertainties, the GRIN-PC lens focuses the wave energy as expected from the gradient refractive index distribution. 

\section{Discussion and Conclusion}
  In this work, we present the conformal GRIN-PC theory based on the ray trajectories in curved GRIN-PC lenses and demonstrate its validity for accurately predicting the focal region of a GRIN-PC lens integrated over a steel pipe. The ray trajectories represent guided wave propagation inside a GRIN-PC lens due to the gradient distribution of the refractive index that results in the focusing of elastic waves. On the other hand, the optical GRIN theory predicts that the ray trajectories in the GRIN-PC lens with hyperbolic secant refractive index distribution meet at a single location without accounting for the crystal anisotropy of phononic crystals. Thus, the predicted focal spot does not agree well with the numerical simulations. The non-planar GRIN theory proposed in this paper utilizes the EFCs of phononic crystal to capture the crystal anisotropy and predict the ray path inside a GRIN-PC lens. The ray trajectories obtained using conformal GRIN-PC theory intersect at multiple locations, and the theoretical focal regions of the GRIN-PC lens are determined by marking the intersection of ray trajectories at the centerline of the lens. The theoretical focal region determined using conformal GRIN-PC theory for L(0,2) pipe mode is in excellent agreement with the focal region obtained in numerical simulations. Thus, the non-planar GRIN theory accurately predicts the entire focal region of a curved GRIN-PC lens overcoming the limitations of optical GRIN theory. Next, to demonstrate the effectiveness of the proposed theory, we present a 3D-printed conical GRIN-PC lens design for multimode focusing of guided elastic waves. For guided wave propagation along the cone axis, the GRIN-PC lens design for conical structure requires uniform angular spacing between the neighboring unit cell rows to respect the cone geometry. This results in varying arc lengths of the curved unit cells as the diameter changes along the cone axis. Thus, the dispersion curves are different for every unit cell, even in the same row of the GRIN-PC lens. As a result, the refractive index profile changes along the cone axis, because of which the conventional GRIN theory fails to predict the focal spot of a conical GRIN-PC lens. We successfully demonstrated the wave-focusing ability of the designed GRIN-PC lens for the three fundamental pipe modes at multiple frequencies through numerical simulations. The non-planar GRIN-PC theory based on the ray tracing framework enables new lens designs conforming or integrated with non-planar geometries and predicts the wave behavior and focal spots in an accurate manner. Thus, it expands the applicability of wave focusing phenomena in a myriad of real-life structures in mechanical, aerospace, and civil engineering applications.

\section*{Data availability statement}
Data is available on reasonable request from the corresponding author.

\section*{Acknowledgements}

This work was supported in part by the National Science Foundation [grant number CMMI-1914583].

\section*{Author contribution statement}

Danawe: Conceptualization. Methodology. Software. Experiments. Validation. Writing- Original draft preparation.  Tol: Conceptualization. Supervision. Writing- Reviewing and Editing. 

\pagebreak
\pagestyle{empty}
\newcommand{\newblock}{}
\bibliographystyle{model1-num-names}

\vskip3pt

\end{document}